\documentclass[12pt,preprint]{aastex}

\newcommand{\mm}{{\rm \, mm}}
\newcommand{\km}{{\rm \, km}}
\newcommand{\yr}{{\rm \, yrs}}
\newcommand{\au}{{\rm \, AU}}
\newcommand{\K}{{\rm \, K}}
\newcommand{\GHz}{{\rm \, GHz}}

\begin{document}

\title{What Can the Cosmic Microwave Background Tell Us About the Outer Solar System?}
 
\author{Daniel Babich\altaffilmark{1,2}, Cullen H. Blake\altaffilmark{1,3} and Charles L. Steinhardt\altaffilmark{1}}
 
\altaffiltext{1}{Harvard-Smithsonian Center for Astrophysics, 60 Garden
St., Cambridge, MA 02138; {\it E-mails}: dbabich,cblake,csteinha@cfa.harvard.edu}

\altaffiltext{2}{California Institute of Technology, Theoretical Astrophysics, MC 130-33,
Pasadena, CA 91125}

\altaffiltext{3}{Harvard Origins of Life Initiative Fellow}

\begin{abstract}
We discuss two new observational techniques that use observations of the Cosmic Microwave Background (CMB) to place constraints upon the mass, 
distance, and size distribution of small objects in the Kuiper Belt and inner Oort Cloud, collectively known as Trans-Neptunian 
Objects (TNOs). The first new technique considers the spectral distortion of the isotropic, or monopole, CMB by TNOs that have 
been heated by solar radiation to temperatures above that of the CMB.  We apply this technique to the spectral measurements of 
the CMB by the Far Infrared Absolute Spectrophotometer (FIRAS) on the Cosmic Background Explorer (COBE).  The second technique 
utilizes the change in amplitude of the TNO signal due to the orbital motion of the observer to separate the TNO signal from 
the invariant extra-galactic CMB and construct a map of the mass distribution in the outer Solar System. We estimate the ability 
of future CMB experiments to create such a map.
\end{abstract}

\keywords{cosmic microwave background -- Kuiper Belt -- Oort cloud}

\maketitle

\section{Introduction}

While we know that in some manner dynamical instabilities in the proto-planetary disk led to the formation of planetesimals 
and planets, the precise details of this process are not well understood \citep{Goldreich04,Lissauer2005}. 
Luckily, just as nature has provided us with the Cosmic 
Microwave Background (CMB) as a relic of the early universe, there are similar relics of this chaotic period of formation.  
Objects in the outer Solar System, collectively known as Trans-Neptunian Objects (TNOs), with semi-major axes 
$40\au \le a \le 10^5 \au$ are believed to provide clues that may help us understand the early history of the Solar System. 
The number density and mass distribution of TNOs contain valuable information about the properties of the proto-planetary 
disk out of which the Solar System formed and the dynamical properties of these TNOs can constrain migratory motion of the 
planets. We show in this paper that the CMB itself may be the key to unlocking the mysteries of these objects.

To date, the entirety of our observational evidence about the structure of the 
Solar System beyond Neptune consists of observations of comets that have been perturbed into the inner 
Solar System and of relatively large, nearby objects detected through their reflected sunlight \citep{Bernstein04}. 
Because of the strong dependence of the reflected sunlight on the object's distance and radius ($\propto D^{-4}, \propto R^{-2}$), 
it is very difficult to find objects smaller or more distant than, for example, (90377) Sedna \citep{Brown04}, a body 
$1180-1800 \km$ in diameter detected at $90 \au$, or 2003 UB$_{313}$ \citep{Brown05}, a body $2400 \km$ in diameter detected near 
its aphelion distance of $96 \au$. 

Long Period Comets (LPC) associated with the Oort Cloud originate in its outermost reaches because objects with large 
semi-major axes can be 
preferentially ejected into the inner Solar System\footnote{Perturbations by Jupiter are large enough to either eject the new 
LPC on a hyperbolic 
orbit or keep it within $1000 \au$ of the Sun.  This originally led Oort to postulate that comets on orbits with large aphelia 
are always entering the inner Solar System for the first time.}. The observed aphelia of LPCs place the outer limits of the 
Oort Cloud at a distance of approximately $25000 \au$ \citep{Oort50,Marsden71}, while simulations of the Oort Cloud's formation, 
assuming that the Sun formed in a star cluster, suggest that a significant amount of mass may lie in an inner Oort Cloud, 
which is located at $1000 \au$ \citep{Hills81,Fernandez97,Dones04}. To this point, we have no methods capable of detecting 
smaller objects in the Kuiper Belt or any objects in the inner Oort Cloud.

In this paper, we develop two new methods capable of exploring the distribution of TNOs.  Both utilize the CMB as a standard 
spectral template, namely a blackbody with a given temperature emanating from a surface at infinity.  The Sun heats Kuiper Belt 
and Oort Cloud objects to a temperature above that of the CMB. These objects extinguish part of the CMB and emit blackbody radiation, 
inducing a spectral distortion in the observed CMB.  By constraining these spectral distortions we can constrain the radial and 
mass distribution of trans-Neptunian objects. 

Some similar topics have previously been discussed.  Detections of zodiacal light, both through its reflected optical sunlight 
and its emitted infrared radiation, have led to detailed models of inter-planetary dust in the Solar System \citep{Fixsen02a}. 
This dust is believed to come from the comae of comets and the collisional debris of asteroids. The infrared emission of Kuiper 
Belt objects has been employed in a similar fashion to this paper in order to constrain the Kuiper Belt \citep{Kenyon01,Teplitz99,
Backman95}.The main advantage of using the CMB as the source of background radiation is that its properties are much better understood 
and more uniform than the far-infrared background, allowing for a greater precision. Additionally, the longer relevant wavelengths of 
the CMB spectrum enable us to apply the same test to the inner Oort Cloud, which is at too low of a temperature to emit any 
substantial amount of radiation in the infrared bands.

In \S~\ref{sec:basics}, we calculate the blackbody radiation emitted by objects in the outer Solar System including a discussion of 
the temperature of distant objects. In \S~\ref{sec:mean}, we use FIRAS to place constraints on the total mass and size distribution 
of the smallest objects in the Kuiper belt. In \S~\ref{sec:WMAP}, we propose a new method of using WMAP to develop a two-dimensional 
map of the smallest objects in the Kuiper Belt, holding out the possibility that for sufficient mass we may be able to determine 
the distance in addition to the mass and direction.  

\section{Outer Solar System Spectral Distortions}
\label{sec:basics}

The extremely high interaction rates between matter and radiation during the early universe cause the primordial 
CMB spectrum to have a blackbody distribution to incredibly high precision \citep{mather1992,Peebles93}. However, as the 
universe expands and these interaction rates decrease, new sources of energy injection can distort the CMB spectrum 
and photon-baryon plasma is now incapable of producing a blackbody spectrum again. In addition to energy injection at 
high redshifts, there are several low redshift processes that are capable of distorting the CMB spectrum 
(see \citet{Tegmark95} and \citet{Fixsen02} for an overview). In this paper we will try to constrain the properties of 
TNOs from the {\it absence} of spectral distortions, so any additional process that may produce these distortions does 
not affect our ability to place these constraints on the outer solar System, but in principle strengthens our conclusions. 
We will consequently ignore these other effects in this paper.

It should be emphasized that these spectral distortions are not the same as the chemical potential ($\mu$) or Compton-$y$
distortions often discussed in the literature \citep{Tegmark95,Fixsen97}. Of course, there is a level of degeneracy
between the spectral distortion considered in this paper, namely a weighted sum of blackbodies, and the chemical potential
and Compton-$y$ distortions, so our signal will result in a non-zero chemical potential and Compton-$y$ distortion if either
of these models was assumed in the analysis.

\subsection{Calculation}

We are interested in the aggregate blackbody emission from objects at a given distance in the outer Solar System. 
The distorted CMB intensity\footnote{There should be no measurable 
polarization signal as the mean CMB and the TNO emission are both 
blackbody radiation. Any polarization signal produced by features 
in the surface of the TNOs will vanish because the experiment's beam 
will contain many TNOs that are randomly oriented with respect to one 
another. If the TNOs are aspherical and have a large magnetic 
susceptibility it is possible that they may align with the magnetic field
via the Greenstein-Davis effect\citep{Draine03}. An analysis of this 
possibility is beyond the scope of this paper.} 
spectrum due to these objects will be
\begin{equation}\label{eq:spec}
\delta I_{\nu} = \tau [B_{\nu}(T_{\rm TNO}) - B_{\nu}(T_{\rm CMB})],
\end{equation}
where the optical depth, or equivalently the geometric covering fraction since we are working in 
the geometric optics limit, is expressed as
\begin{equation}\label{eq:tau}
\tau = \int \frac{\pi R^2}{4\pi D^2} n(M) dM,
\end{equation}
where $R$ is the radius of the object and $D$ is the approximately identical distance to the TNO from both the Sun and 
the observer. These two distances are not exactly equal and the TNO-observer distance may be time-variable. We will explore
this possibility in \S \ref{sec:WMAP}.

In this paper we assume that all of the TNOs are at a single, unknown distance which is then constrained by
observations. Of course, TNOs can have quite different eccentricities and semi-major axes so at a given time there
will be a radial distribution of TNOs. Since both the geometric covering fraction and TNO temperature decrease
with increasing distance from the Sun, we will be most sensitive to the closest TNOs. This helps to justify our
assumption of a single distance. Moreover, the TNO temperature varies with distance from the Sun so the radial 
distribution of TNOs will produce a more complicate spectral distortion than implied by Eq. (\ref{eq:spec}). With 
good enough data this radial distribution could be constrained by constraining the distribution of TNO temperatures. 
However, with the limited sensitivity of current data our assumption of a single distance is sufficiently accurate.
In \S \ref{sec:WMAP} we will slightly relax this assumption by allowing the TNO distances to be anisotropic; therefore,
all of the TNOs in a given pixel will be at the same distance, which will be different than the TNO distance in
other pixels. 

The size distribution of TNOs is specified by their mass function, $n(M)$, with total mass
\begin{equation}
M_{\rm total} = \int n(M) M dM. 
\end{equation}

For Kuiper Belt objects, the heating is almost entirely solar, so we can calculate the equilibrium temperature of 
these objects. In the case of Oort Cloud objects, their equilibrium temperature even at large distances is higher 
than that of the CMB, even though at 25000 AU the solar contribution is tiny. Interstellar processes, as well as
the absorption of CMB photons, results in a floor for the TNO temperature. The equilibrium temperature of Oort 
Cloud objects is assumed not to drop below 5-6 K, as claimed by \citet{Mumma93} and \citet{Stern2003}. In 
this work we will calculate the temperature of TNOs by assuming 
thermal equilibrium with the Sun. As mentioned above, this assumption breaks down for TNOs with extremely large semi-major 
axes ($a \ge 10000 \au$), but objects at such a large distance will produce a signal well below current instrumental 
sensitivities and therefore are irrelevant for our work. Assuming thermal equilibrium with the Sun, the temperature of a TNO is
\begin{equation}\label{eq:temp}
T_{\rm TNO} = \left[\frac{R^2_{\odot}}{4 D^2}(1-A)\right]^{1/4} T_{\odot},
\end{equation}
where $R_{\odot}$ is the Sun's radius, $T_{\odot}$ the temperature of the Sun's photosphere, and $A$ the
TNO's albedo. At distances appropriate for the Kuiper Belt, $D = 40 \au$, we find a temperature of $T = 43 \K$
and at inner Oort Cloud distances, $D = 1000 \au$, the temperature is $T = 9 \K$.

The total distortion will depend on the following parameters of the Oort Cloud and Kuiper Belt. Here we 
describe our choices for these parameters, as well as theoretical and observational constraints.

\begin{itemize}
\item Density - We assume a density $\rho = 1 \mbox{ g cm}^{-3}$. This number depends on the porosity of the TNOs 
and may vary by a factor of a few.

\item Albedo - We will assume an albedo of $A = 4 \%$ as has been measured for comets and is relevant for 
dirty ice \citep{Luu02}. 
The constraints only depend on the albedo as $(1-A)^{1/4}$ so this uncertainty does not strongly affect our results.

\item Mass Function - We will assume a broken power law\footnote{Typically the differential mass function is
expressed in terms of comet radius, not mass. The power law with respect to radius is related to the values 
used in this paper as $\alpha_{\rm mass} = (\alpha_{\rm radius}+2)/3$. The canonical faint-end power-law exponent of $\alpha_{radius}=3.5$ implies $\alpha_{mass}=1.83$. }

\begin{eqnarray*}
n(M) = & A M^{-\alpha}, & M_{\rm min} < M < M_{\rm br} \\
n(M) = & A M_{\rm br}^{-\alpha+\beta} M^{-\beta}, & M_{\rm br} < M < M_{\rm max},
\end{eqnarray*}
The appropriate power law exponents are not well known. We will find that our results are quite sensitive to
$\alpha$, the low mass slope, because for a given total amount of mass in either the Oort Cloud or Kuiper
Belt a steeper mass function will increase the geometric covering fraction. It is difficult to observationally 
constrain the mass function of comets due to the uncertainties in the dynamics of comet's comae. Collisional 
equilibrium, along with the assumption that the strength of the object is independent of size, would lead us 
to expect $\alpha = 11/6$ \citep{Pan05}. If the larger objects are more difficult to shatter, as would occur if the 
gravitationally binding energy were very important, then $\alpha < 11/6$. Since there is so much uncertainty 
in $\alpha$ we will present results for several values. The value of $\beta$ has little effect upon our final 
results if $\beta > 2$. We will adopt a single value of $\beta = 13/6$, which is consistent with the current
observational results \citep{Bernstein04,Pan05}. In \S \ref{sec:concl} we will discuss how our conclusions depend on
this assumption.

\item Lower Mass - The minimum size of surviving objects is determined by the ability of Robertson-Poynting 
drag to eliminate the smallest objects as well as the properties of collisions that can fragment larger
objects into smaller ones \citep{Burns79}. We will take this lower mass to correspond to objects of radius
$1 \mm$. This quantity depends on the detailed formation history of the Kuiper Belt and the Oort Cloud and 
is difficult to calculate. This quantity is partially degenerate with the total mass of the outer 
Solar System, which is also unknown, so for the purposes of this paper we will fix it at $1 \mm$.

However, we should note that the efficiency of blackbody emission is suppressed when the radiation wavelength 
is larger than the size of the object emitting the radiation due to Kirchoff's Law \citep{Greenberg78}. This suppression 
of the TNO emissivity, which will change the low frequency portion of the CMB spectrum, may be used to constrain the 
minimum mass. We will discuss this in more detail in the \S \ref{sec:mean} where we mention a new proposed CMB
experiment to measure these low frequencies.

\item Upper Mass - We assume $M_{\rm max} = 10^{-2} \times M_{\rm Earth} = 6 \times 10^{25} \mbox{ g}$ as the upper mass.  
Our results are insensitive to changes of even many orders of magnitude in the upper 
mass limit for $\beta > 2$. If $\beta < 2$, a significant fraction of the mass would be at the high-mass 
end and our limits on the low-mass end would not directly translate to limits on the overall mass in 
the outer solar system.

\item Break Mass - $M_{\rm br} = 3.2 \times 10^{19} \mbox{ g}$ (corresponding to $R_{\rm br} = 2 \times 10^6 \mbox{ cm}$). 
The break mass is analytically calculated by determining the largest TNOs that can be in collisional equilibrium over
the lifetime of the Solar System \citep{Pan05}. Like the power law indices, this parameter is also uncertain but, like 
the upper mass, our results are relatively insensitive to its value. Observations with the Hubble Space
Telescope have constrained the break radius to be $R_{\rm br} \le 20 \rm{ km}$ \citep{Bernstein04}.

\item Total Mass - The total mass $M_{\rm tot}$ is one of the things which we seek to constrain with this 
calculation. Theoretically the total mass should be between $(0.1-100) \times M_{\rm Earth}$ (cf. \citet{Stern2001}), 
depending on the surface mass density of proto-planetary disk and the details of the growth of the planets. 

\item Distance - We consider what limits can be set for distances from $D = 40 \au$ to 
$D = 10000 \au$. The strength of the effect roughly scales as $D^{-4}$ so we are only sensitive 
to the inner Oort Cloud and not the outer Oort Cloud from where LPC are observed to originate.

\end{itemize}

\subsection{Strategies}

In this subsection we will outline the two strategies that can be used to constrain the mass distribution in
the outer Solar System via their spectral distortions to the CMB. For small optical depth $\tau$, the observed 
intensity fluctuation in the CMB towards a given direction $\hat{n}$ is
\begin{eqnarray}\label{eq:Int}
I_{\nu}(\hat{n}) = [1-\tau(\hat{n})]\left[B_{\nu}(\bar{T}_{\rm CMB}+ \Delta T(\hat{n}))\right] \\ 
+ \hat{n}\cdot\vec{v} + \tau(\hat{n}) B_\nu(T_{Oort}(\hat{n})) + N_{\rm Instr}, \nonumber
\end{eqnarray}
corresponding to the extragalactic CMB, including the temperature anisotropy ($\Delta T(\hat{n}))$), 
extinguished by the TNOs in the beam; the Doppler effect, assuming that the observer has velocity $\vec{v}$ 
with respect to the CMB restframe; TNO blackbody emission; and instrument noise, which is statistically 
stationary but random. In what follows, we will assume that the
anisotropy produced via the Doppler effect, while in principle large, is calculable and therefore does not
affect our ability to extract or constrain the TNO component of the signal. Subtracting off the mean CMB 
blackbody, the intensity fluctuation in a given direction is 
\begin{eqnarray}\label{eq:dInt}
\delta I_{\nu}(\hat{n}) &=& \frac{\partial B_{\nu}(\bar{T}_{\rm CMB})}{\partial T} \Delta T(\hat{n}) 
+ \tau(\hat{n}) \\
&\times&\left[B_\nu(T_{Oort}(\hat{n})) - B_{\nu}(\bar{T}_{\rm CMB} + \Delta T(\hat{n}))\right]  
+ N_{\rm Instr}, \nonumber
\end{eqnarray}
Unfortunately for our purposes, the CMB has small statistical temperature fluctuations\footnote{This implies 
that it is impossible to predict the exact pattern of observed temperature anisotropies. We can only predict 
the two-point correlation function (the scale-dependent variance) which parameterize the probability distribution 
function of which the observed CMB is a random realization.} on the order of 1 part in $10^5$. 
We would like to constrain the TNO optical depth, $\tau(\hat{n})$, in a given direction. This is complicated
by the random terms in Eq. (\ref{eq:dInt}) corresponding to instrument noise and CMB temperature anisotropies.

We should be able to directly distinguish the $\ge 5 \K$ blackbody spectrum from the CMB blackbody spectrum,
especially when we include data at higher frequencies ($\nu > 200 \GHz$), at a level set by either the CMB 
temperature anisotropies or instrument noise. This would enable us to directly make a map of the TNO mass density.
If the instrument pixel noise dominates both the intrinsic variation due to the CMB temperature anisotropies and
signal from the outer Solar System, then the constructed TNO map would have a signal-to-noise ratio of less than one
and would be useless. In this case, we can reduce the instrument noise by smoothing the CMB map and averaging together 
neighboring pixels; this will increase the pixel signal-to-noise ratio. The extreme limit of the procedure is to 
calculate the mean CMB spectrum. 
If the CMB temperature anisotropies dominate the instrument noise it  is possible  to reduce, 
but not completely eliminate, the variance of the temperature anisotropies. This technique is based on differencing 
the observations of a given position of the Celestial sphere made at different locations within the Solar System. It 
will be discussed in detail in \S \ref{sec:WMAP}. Using this technique it may be possible to make a two- or even three-, 
dimensional map of the distribution of TNOs.

\section{Isotropic Spectral Distortion}
\label{sec:mean}

\subsection{Observational Constraints}

We use data from the {\it Far Infrared Absolute Spectrophotometer} (FIRAS) on the Cosmic Background Explorer (COBE) \citep{Fixsen97}
to place limits on deviations from a blackbody spectrum due to TNOs. FIRAS measured the mean, or monopole, CMB 
frequency spectrum so we will average the signal originating in the outer Solar System over the entire sky in order to
compare with their results. The Oort Cloud is assumed to be isotropic since the observed distribution of 
inclination angles of new LPCs is approximately uniform \citep{Marsden71}, although the Galactic tide \citep{Heisler86} 
and stellar perturbations \citep{Babich06} should make it aspherical. 

The Kuiper Belt is highly anisotropic, since it is primarily located near the ecliptic plane. The FIRAS results we
use have been averaged over the full-sky in each frequency band and therefore the signal from the Kuiper Belt 
must also be averaged over the full sky. Since we know that the Kuiper Belt is located at small inclination,
it is less optimal to use the full sky averaged FIRAS spectrum. This will still allow us to place constraints
on the Kuiper Belt, even if they are not the best possible.

Since FIRAS did not detect any spectral distortion in the CMB we can only place upper limits on the basic 
parameters of the outer Solar System. There are a multitude of processes that can produce 
spectral distortions but we will ignore them as we are simply using the lack of detection to 
constrain the properties of the outer Solar System. Note that the inclusion of these other effects would 
only make our upper limits stronger.

\begin{figure}[b!]
\centerline{
\includegraphics[width=\textwidth]{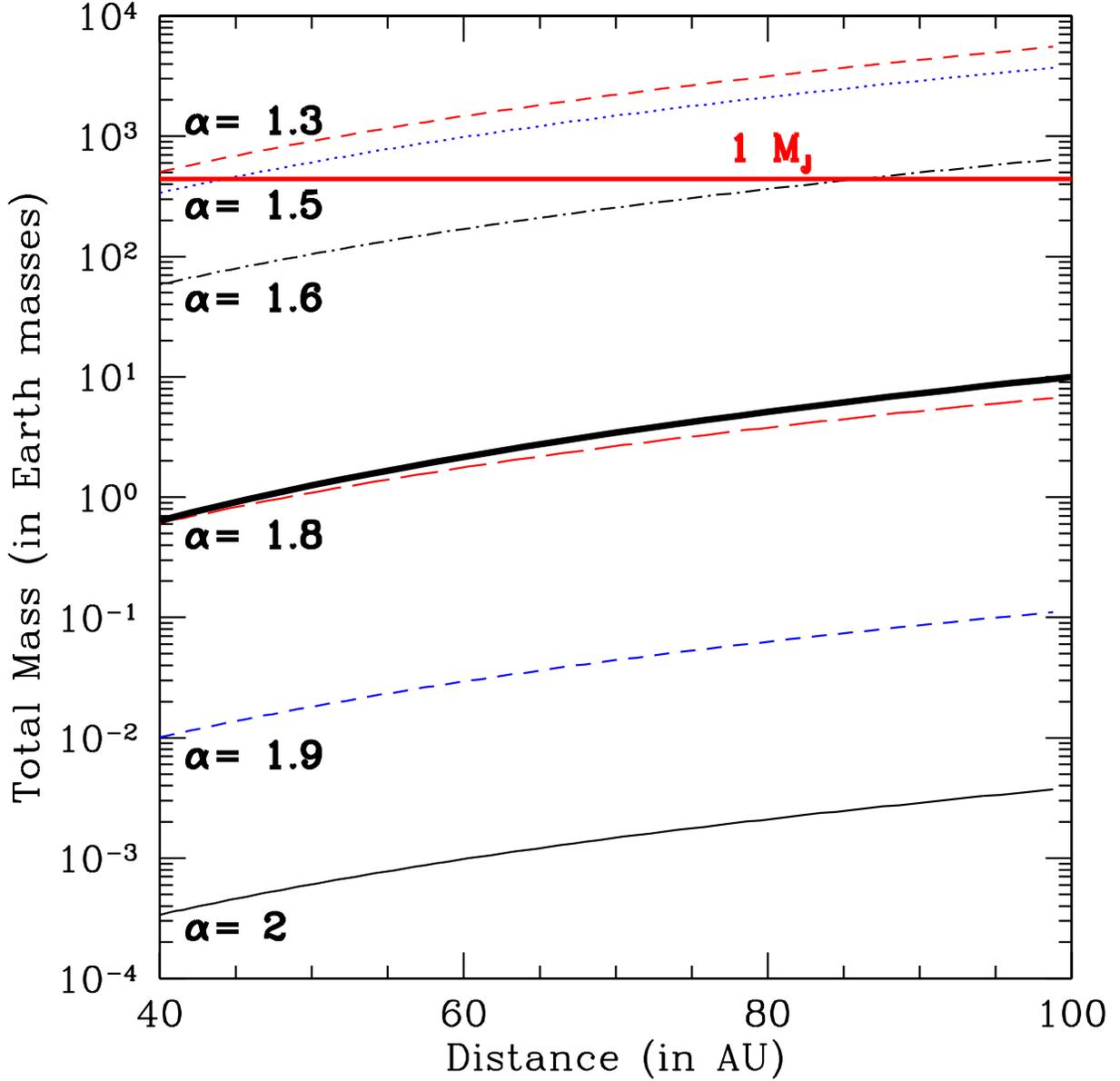}}
\caption{\label{fig:kb}
Excluded regions of total mass and distance for the Kuiper Belt for different mass function slopes $\alpha$.
The curves correspond to $\alpha = 2$ (black, solid); $\alpha=1.9$ (blue, dashed); $\alpha=1.8$ (red, long-dashed); 
$\alpha = 1.667$ (black, dot-dashed); $\alpha = 1.5$ (blue, dotted) and $\alpha = 1.333$ (red,dashed). 
The region above the curves is excluded at $95\%$ by the COBE FIRAS data. Also shown is a dynamical 
limit determined by constraints on perturbations to the orbit of Halley's comet (heavy black, solid curve)
and, for reference, a line corresponding to a Jupiter mass (heavy red, horizontal line). The power law with respect to radius is related to the values used in this paper as $\alpha_{mass}=(\alpha_{radius}+2)/3.$}
\end{figure}

\begin{figure}[b!]
\centerline{
\includegraphics[width=\textwidth]{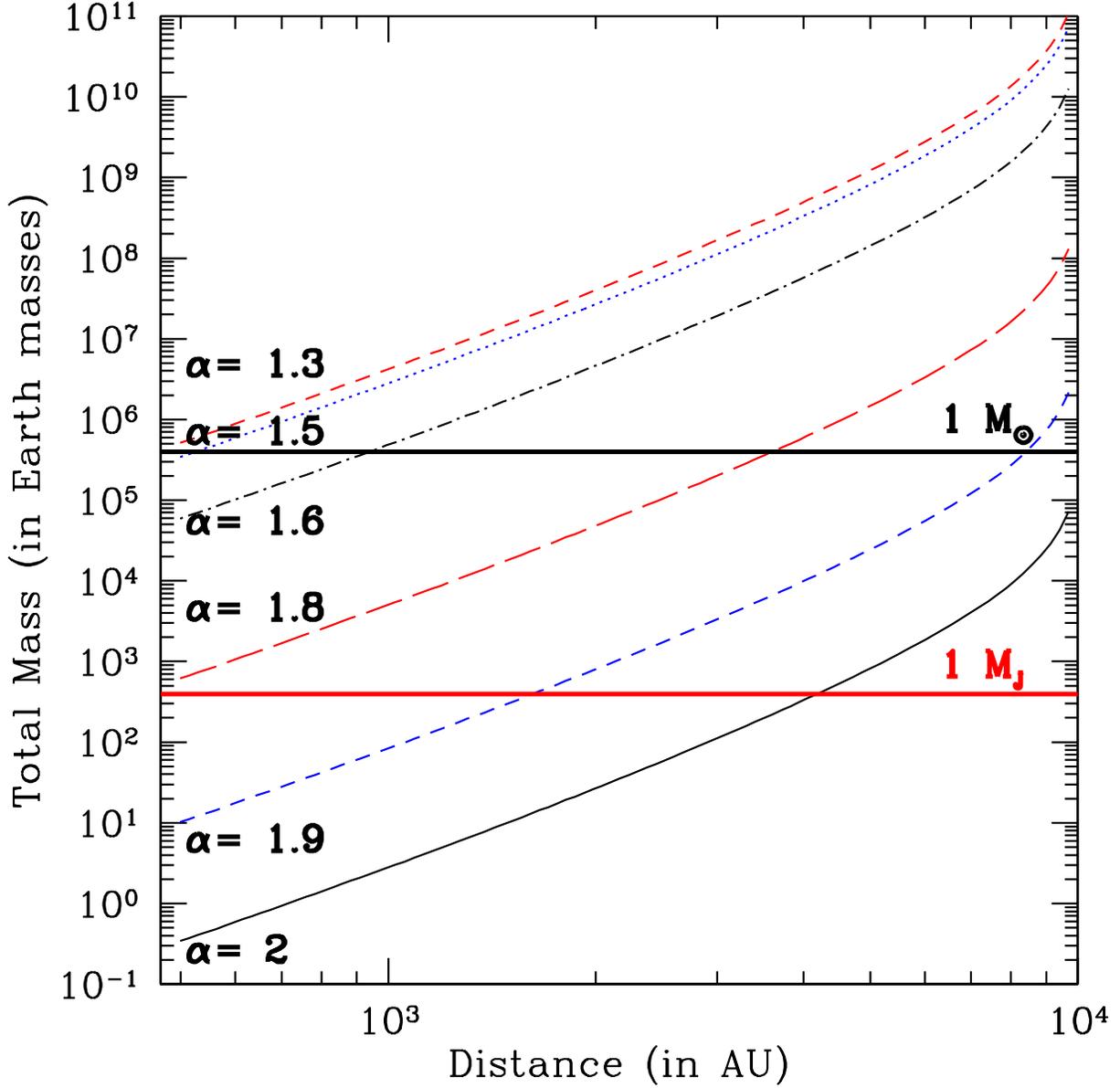}}
\caption{\label{fig:oort}
Excluded regions of total mass and distance for the inner Oort Cloud for different mass function slopes $\alpha$.
The curves correspond to $\alpha = 2$ (black, solid); $\alpha=1.9$ (blue, dashed); $\alpha=1.8$ (red, long-dashed); 
$\alpha = 1.667$ (black, dot-dashed); $\alpha = 1.5$ (blue, dotted) and $\alpha = 1.333$ (red,dashed). 
The region to the upper left of the curves is excluded at $95\%$ by the COBE FIRAS data. For reference we also
include lines corresponding to a solar mass (heavy black, horizontal line) and a Jupiter 
mass (heavy red, horizontal line).}
\end{figure}

Figures \ref{fig:kb} \& \ref{fig:oort} display constraints on the total mass\footnote{Our technique is 
mainly sensitive to low mass objects so we must be very cautious when interpreting these results. First of 
all, we are assuming that the mass function is well modeled as a broken power-law over a very wide range 
in mass. Second, Poisson fluctuations become important at the high mass end, where the expected number
of objects, as extrapolated from the mass function, approaches unity. Additionally, run-away growth during
the core accretion phase of formation can grow a single object to a much larger mass than any other object
in its vicinity. A simple extrapolation of the mass function will miss these objects. These last two effects 
can cause the actual mass of the Kuiper Belt or Oort Cloud to differ significantly from our constraints. 
Fortunately, these high mass objects are more easily detected in optical surveys and so our technique is 
complimentary to existing methods.} 
and the distance of the Kuiper Belt and inner Oort Cloud, respectively, from a non-detection of CMB 
spectral distortions in the FIRAS data. The confidence contours are determined by calculating the change 
in the $\chi^2$ of the FIRAS data when a component originating from the outer Solar System is included.
The errors used to calculate the $\chi^2$ are taken from Table 4 of \citep{Fixsen96} for the low 
frequency FIRAS data which extends up to $630 \GHz$.

The curves are shown for several values of $\alpha$ since 
the constraints strongly depend on the low mass end of the planetesimal mass function. Even though the 
Kuiper Belt primarily lies in the ecliptic plane and the FIRAS data is the mean CMB spectrum, 
averaged over the entire sky, we can still use this data to constrain, in a sub-optimal manner, 
the properties of the Kuiper Belt. These constraints correspond to $95\%$ confidence limits.
Also shown is a dynamical constraint on the total mass in the Kuiper Belt from constraints
on the orbit of Halley's comet \citep{Hogg91,Hamid68}. 
These figures demonstrate that our technique is competitive with dynamical limits on the Kuiper
Belt mass and may place more stringent constraints if the slope of the low mass end of the mass function 
is steep enough. Hence, if $\alpha_{mass}=3.5$ ($\alpha_{radius}=1.83$) then the dynamical CMB results place limits to the TNO mass that are as tight as the dynamical limits from Halley's comet's orbit, and exclude $M>1M_{\sun}$ in general. This technique may be the only way we can constrain the properties of the inner
Oort Cloud.
Of course, our technique is only sensitive to the low mass end of the distribution
and so care must be taken when interpreting these results.

\clearpage


\clearpage

\subsection{Low Frequency Observations}
A new experiment, the {\it Absolute Radiometer for Cosmology Astrophysics and Diffuse Emission} (ARCADE), has been 
proposed to constrain low frequency CMB spectral distortions \citep{Kogut06}. This experiment would measure the 
CMB spectrum over the full sky in narrow bands between $3 \GHz$ and $30 \GHz$. The addition of these lower frequencies
to the COBE FIRAS data would significantly help in constraining the mean Compton-$y$ parameter and chemical potential.
If the TNOs emit as perfect blackbodies, the low frequency data would not help us constrain objects in the outer Solar System.
At these low frequencies, where both the TNO emission spectrum and the CMB spectrum are both in the Rayleigh-Jeans 
limit we would simply measure a spectrum also in the Rayleigh-Jeans limit at the composite temperature
\begin{equation}
T = (1-\tau)T_{\rm CMB} + \tau T_{\rm TNO}.
\end{equation}
Since the CMB temperature is {\it a priori} unknown it is not possible to exclude the contamination of another 
blackbody at a different temperature using only low frequency observations. Measurements in the Wien portion of 
the spectrum allow us to constrain the properties of the outer Solar System since at these relatively high 
frequencies the CMB and TNO spectra are significantly different. However, for small enough particles, or low 
enough frequencies, the TNO emissivity will deviate from a prefect blackbody \citep{Greenberg78} and the emitted
spectrum will not be well described by the Rayleigh-Jeans formula and the low frequency data will become useful. 

The emissivity, $j(\nu)$, of a body in thermal equilibrium at temperature $T$ is determined by Kirchoff's 
law \citep{Rybicki79}
\begin{equation}
j(\nu) = \alpha(\nu) B_{\nu}(T),
\end{equation}
where $\alpha(\nu)$ is the absorptivity and $B_{\nu}(T)$ is the Planck function. In the geometric optics limit,
which we employed in the previous subsection, the absorptivity is simply determined by the frequency-independent 
geometric cross section. When the particle size becomes comparable to the radiation's wavelength and the
geometric optics limit becomes invalid the absorptivity becomes frequency dependent \citep{Spitzer78}. These 
deviations may be used to constrain the minimum mass of TNOs.

\section{Anisotropic Distance Modulation}
\label{sec:WMAP}

In the previous section we considered the spectral distortion of the mean CMB when averaged over a long
time period. Implicit in that analysis was the assumption that the Sun-TNO distance ($\bar{D}$) and observer-TNO
distance ($D$) are identical and constant. The Sun-TNO distance, $\bar{D}$, which determines the TNO's temperature, 
is constant during the lifetime of observation since the relevant periods are approximately $T \sim 300 \yr$ 
for the Kuiper Belt and $T \sim 31000 \yr$ for the inner Oort Cloud. $D$ would change during the course of the 
observations if the satellite looks at the same point on the Celestial sphere from different locations within the 
Solar System. This depends on the detailed scan pattern of the relevant experiment. WMAP does view the same 
position on the Celestial sphere from different positions but Planck will not.

If the scan pattern of the relevant experiment is such that it does observe the same point of the Celestial sphere
from multiple distances, then the time-independent extragalactic CMB signal can be partially removed and the
component arising from the TNO can be better constrained. In this section we study this technique by considering the
various time varying signals and calculating potential constraints that could be produced.

\subsection{Distance-Dependent Signal}
The observed intensity fluctuation in the CMB towards a given direction $\hat{n}$, Eq.(\ref{eq:Int}), is 
dependent on the TNO-observer distance as
\begin{eqnarray}
I_{\nu}(\hat{n},D) = [1-\tau(\hat{n},D)]\left[B_{\nu}(\bar{T}_{\rm CMB}+ \Delta T(\hat{n}))\right] \\ 
+ \hat{n}\cdot\vec{v}(D) + \tau(\hat{n},D) B_\nu(T_{TNO}(\hat{n}),\bar{D}) + N_{\rm Instr}. \nonumber
\end{eqnarray}

Attempting to pick out the Oort Cloud blackbody directly is difficult due to the statistical nature of the 
temperature anisotropies, especially if we are restricted to low frequency data. However, by looking at the 
same position on the Celestial sphere from different points in the CMB experiment's orbit, we can alter the 
distance between the observer and the TNO. This difference is
\begin{eqnarray}\label{Int_Diff}
I_{\nu}(\hat{n},D_i)&-&I_{\nu}(\hat{n},D_j) = [\tau(\hat{n},D_i) - \tau(\hat{n},D_j)] \nonumber \\
&\times&[B_\nu(T_{TNO}(\hat{n}),\bar{D}) - B_{\nu}(\bar{T}_{\rm CMB} + \Delta T(\hat{n}))] \nonumber \\
&+&\hat{n} \cdot [\vec{v}(D_i) - \vec{v}(D_j)] + N^i_{\rm Instr} - N^j_{\rm Instr}.
\end{eqnarray}
By taking this difference we have reduced the importance of the portion of the signal 
($\Delta T \times \partial B_{\nu}(\bar{T}_{\rm CMB})/\partial T$) that is statistically unknown. 
Other terms are assumed to be known to such a precision that they be cleanly removed.  So, we consider each change 
in the intensity fluctuation accompanying a change in the CMB, as our detection limits for the TNO are set by the 
largest accompanying change.

\subsubsection{Constancy of the CMB} \label{sec:const}
The CMB anisotropies as measured from two different locations are not necessarily
identical. There are three effects that potentially could cause small changes: {\it (i)} each set of anisotropies have 
slightly different surfaces of last scattering and therefore depend on slightly different initial curvature perturbations; 
{\it (ii)} the path difference between the two different observation points can induce phase shifts in the anisotropy 
Fourier coefficients, $a_{\ell m}$, due to free streaming of the anisotropies and {\it (iii)} different ordinary Sachs-Wolfe 
effect due to different values of the solar system gravitational potential at these different points. The first two effects
produce very small changes and are negligible. Note that the ordinary Sachs-Wolfe effect produced by the Sun can be large:
\begin{equation}
\frac{\Delta T}{\bar{T}_{\rm CMB}} = \frac{2 G M_{\odot}}{c^2 (1 \au)} \sim 1.89 \times 10^{-8};
\end{equation}
however, the uncertainty introduced by the effect is suppressed by the fractional uncertainty in
the distance between the CMB experiment and the Sun. The constancy of the CMB anisotropies 
only allows us to reduce the influence of the anisotropies and their statistical uncertainty from
$\mathcal{O}(\Delta T)$ to $\mathcal{O}(\tau \Delta T)$, not completely eliminate them. 

\subsubsection{Doppler Effect}  The relative velocity of the observer with respect to the CMB rest frame
induces anisotropies into the observed CMB. For small velocities the Doppler effect can be decomposed into 
contributions from the motion of the Solar System with respect to the CMB rest frame and the motion of the 
observer around the Sun\footnote{The motion of the observer around the Sun will also induce a Doppler shift 
in the radiation emitted by the TNOs. This effect is smaller than the Doppler shift of the extragalactic 
CMB by a factor of the TNO optical depth, $\tau$.}. The first velocity, the motion of the Sun with respect 
to the CMB restframe, will be constant on the time scale of observations. As we move to a different 
position in the Earth's orbit the Doppler contribution $\hat{n}\cdot \vec{v}$ will also change and the 
two Doppler effect terms in Eq. (\ref{Int_Diff}) will not in general cancel. Even though the anisotropy 
produced by the Doppler effect is large, its uncertainty is not. The uncertainty in the Earth's orbit is 
of order $10^{-11}$ in position and velocity and the uncertainty in the Doppler effect contribution will 
be of the same order \citep{Standish2004}.

\subsubsection{Trans-Neptunian Objects}  
The temperature of a given TNO is unknown (since the distance to the object is unknown), but is assumed to be constant
during the period of CMB observations. The fraction of our beam filled by TNOs depends on the variable observer-TNO
distance. For all reasonable choices of Kuiper Belt and inner Oort Cloud parameters the beam is sparsely filled
and therefore the TNO filling fraction will vary as $D^{-2}$.\footnote{If the beam were completely filled then the
distance modulation would not change the observed surface brightness.}

Some fraction of the TNOs in our beam might be shifted out of the beam due to the parallax effect (especially 
when we are considering the much closer Kuiper Belt) as the CMB experiment changes its position within the Solar System.
The fractional change of the beam filling fraction due to the parallax is
\begin{eqnarray}
f &=& \frac{2 \pi \theta_B \Delta \theta}{\pi \theta^2_B},
\end{eqnarray}
where $\theta_B$ is the beam width and $\Delta \theta = 2 AU/ D$ is the maximum parallax angle in the small angle 
approximation. The need to minimize this fractional change will determine the resolution of the TNO map that we can 
ultimately reconstruct. By requiring this fractional change to be less than $1\%$ we find
\begin{equation}
\theta_B \ge 200 \frac{1 \au}{D}.
\end{equation}
For the Inner Oort Cloud ($D \approx 10^3 \au$) $\theta_B \ge 10'$ and for the Kuiper Belt much larger. 

In principle we can circumvent this parallax problem and construct maps with resolution comparable to the beam
width of the experiment by assuming that the motion of the TNO on the Celestial sphere is completely due to the 
motion of the experiment and then
accounting for the shift between the extragalactic CMB and the TNOs. This is extremely complicated but possible for low 
instrument noise and enough interlocking observations. The details will be presented in a future paper.

When we change the distance to a group of TNOs from $D$ to $D+x$, the fractional change in the TNO 
contribution to the observed CMB intensity is
\begin{equation}
\frac{\delta I}{I} = \frac{D}{2x + x^2/D}.
\end{equation}
The first-order term is a good approximation for $D \gg x$, but for shorter distances the second-order term will also 
contribute.  For $D \gg x$ we will be able to produce a two-dimensional map of the optical depth from small objects using 
this method, but for distances such that the second-order term is significant we will also be able to determine the correct 
distances to these objects and produce a three dimensional map. The number of TNOs in the beam, which is also unknown, is
completely degenerate with the distance $D$ when $D \gg x$ and it is the second order variation that breaks this degeneracy
and allows us to reconstruct both the beam filling fraction and the distance to the TNOs. We consider the potential of
this effect to constrain TNO properties in the next subsection.

\subsubsection{Instrument Noise} 
Each pixel will have a fluctuation due to instrument noise since the system temperature is above the CMB temperature. For 
reference we will describe instrument noise in temperature units ($\sigma_T$) and relate it to uncertainty in the
observed intensity as \citep{Tegmark96}
\begin{equation}
\sigma_B(\nu) = \frac{\partial B_{\nu}(\bar{T}_{\rm CMB})}{\partial T} \sigma_T.
\end{equation}

\subsection{Potential Constraints}
We will now determine how well this technique can produce a map of TNOs for two model experiments, one with 
WMAP instrumental parameters and one with next generation parameters. We will ignore the Doppler and Sachs-Wolfe 
effects discussed in \S \ref{sec:const} and assume that their amplitudes can be determined with sufficient precision 
that we can correct for them.
The Fisher matrix formalism will be employed to estimate these potential constraints
(see Tegmark et al. 1997 for an overview of the Fisher Matrix formalism).

The intensity in a given direction depends on our orbital position because of the portion coming
from the outer Solar System. We can isolate this distance-dependent piece by differencing the observed CMB 
intensities in a given direction as measured at different orbital positions:
\begin{eqnarray}
\Delta \bar{I}_{\nu}(\hat{n}) &=& \sum_{i \ne j} I_{\nu}(\hat{n}, \vec{x}_i) - I_{\nu}(\hat{n}, \vec{x}_j), \\
&=& \sum_{i \ne j} [\tau(\hat{n}, \vec{x}_i)-\tau(\hat{n}, \vec{x}_j)] [B_{\nu}(\bar{T}_{\rm TNO}) 
- B_{\nu}(\bar{T}_{\rm cmb})], \nonumber
\end{eqnarray}
where the optical depth in a given direction is modeled as having an unknown constant amplitude $\tau_0$ in addition to 
the unknown variable distance  
\begin{equation}
  \tau(\hat{n}, \vec{x}) = \frac{\tau_0(\hat{n})}{(\bar{D}(\hat{n}) + x_i)^2}.
\end{equation}
Note that $T_{\rm TNO}(\hat{n})$ also depends on $\bar{D}(\hat{n})$ through the thermal balance 
of Solar heating, Eq. (\ref{eq:temp}).

In order to estimate the precision with which we can measure the distance to an trans-Neptunian object 
we will perform a Fisher matrix analysis. The Fisher matrix is defined as
\begin{equation}\label{eq:fisher}
F_{\alpha \beta} \equiv \sum_{\nu} \frac{1}{\sigma^2_B(\nu)} \frac{\partial \Delta I_{\nu}}{\partial p_{\alpha}} 
\frac{\partial \Delta I_{\nu}}{\partial p_{\beta}}, 
\end{equation}
where $p_{\alpha} = \{ \tau_0(\hat{n}), \bar{D}(\hat{n})\}$.
Since $\tau_0$ is unknown, we marginalize over this parameter, and find the constraints on $\bar{D}$. Uncertainties
in the CMB detector position and velocity, as well as uncertainties in the positions of planets within the Solar
System, will all increase the overall uncertainty.  We expect instrument noise to be the dominant source 
of uncertainty and will assume the noise in Eq. (\ref{eq:fisher}) solely comes from the instrument noise.

\begin{table}
\begin{tabular}{|c|c|c|c|c|c|}
\hline
Mass & Distance & $\alpha$ & $\sigma_T$ & $\sigma_{\tau_0}/\tau_0$ & $\sigma_D/\bar{D}$ \\
\hline
\hline
$1 M_{\Earth}$ & $40 \au$ & $1.5$ & $1 \mu K$ & $19300$ & $6470$ \\
$1 M_{\Earth}$ & $40 \au$ & $1.66$ & $1 \mu K$ & $3280$ & $1100$ \\
$1 M_{\Earth}$ & $40 \au$ & $1.8$ & $1 \mu K$ & $34.6$ & $11.6$ \\
$1 M_{\Earth}$ & $40 \au$ & $1.9$ & $1 \mu K$ & $0.575$ & $0.193$ \\
$1 M_{\Earth}$ & $40 \au$ & $2$   & $1 \mu K$ & $0.0192$ & $0.00647$ \\
\hline
$1 M_{\Earth}$ & $40 \au$ & $1.5$  & $30 \mu K$ & $578000$ & $194000$ \\
$1 M_{\Earth}$ & $40 \au$ & $1.66$ & $30 \mu K$ & $98500$ & $33100$ \\
$1 M_{\Earth}$ & $40 \au$ & $1.8$  & $30 \mu K$ & $1040$ & $349$ \\
$1 M_{\Earth}$ & $40 \au$ & $1.9$  & $30 \mu K$ & $17.2$ & $5.80$ \\
$1 M_{\Earth}$ & $40 \au$ & $2$    & $30 \mu K$ & $0.577$ & $0.194$ \\
\hline\hline
$50 M_{\Earth}$ & $1000 \au$ & $2$ & $1 \mu K $ & $30.0$ & $10.0$ \\
$50 M_{\Earth}$ & $5000 \au$ & $2$ & $1 \mu K $ & $8700$ & $2910$ \\
\hline
\end{tabular}
\caption{\label{table:masses} 
Detection limits of the optical depth, $\tau_0(\hat{n})$, and distance, $\bar{D}(\hat{n})$, for Kuiper Belt 
and Oort Cloud objects. The optical depth and distance are partially degenerate, so the uncertainties presented
here have marginalize over the second degenerate parameter. In order to make a meaningful map of the outer 
Solar System the fractional uncertainties must be less than one.}
\end{table}

These limits allow us to detect the presence of TNOs in a given direction. To first order in 
$\Delta x/\bar{D} \sim 1 \au/\bar{D}$ we are only sensitive to $\tau_0/\bar{D}^5$, so our 
results would be completely degenerate between these two unknown parameters. This degeneracy is 
broken at second order in $\Delta x/\bar{D}$. We can either try to make a two-dimensional map and 
constrain the TNO optical depth in each pixel or we can use second order variations in $\Delta x/\bar{D}$ 
to make a three dimensional map and determine both the optical depth in a pixel as well as the mean 
distance to TNOs in that pixel. Table \ref{table:masses} shows the potential constraints on $\tau_0$ 
and $\bar{D}$ that can be produced by a CMB experiment with a given pixel noise and that observes in ten 
frequency bands and measures a given pixel on the sky with ten different projected distances between 
$-0.5 \au \le \Delta x \le 0.5 \au$. 

The upper portion of the table corresponds to the Kuiper Belt. The plausibility of this technique
strongly depends on the value of $\alpha$. Our results imply that a map of the Kuiper Belt can be made
if $\alpha \approx 2$. In fact, there is a transitional value of $\alpha_{mass} \ge 5/3$ ($\alpha_{radius}  \ge 1.22$) corresponding to
the point where the integral in Eq. (\ref{eq:tau}) becomes dominated by the low mass end of the 
distribution. The lower portion of the table corresponds to the inner Oort Cloud. Due to the strong 
distance dependence of the effect producing a map of the inner Oort Cloud will be much more challenging.
Assuming $\alpha = 1.9$ and $\sigma_T = 1 \mu \K$, there needs to be $M_{\rm total} = 0.185 \times M_{\rm Earth}$ 
in order to produce a two dimensional map with signal-to-noise unity.

\section{Discussion}
\label{sec:concl}

In this paper we describe two new techniques to constrain the cumulative mass contained in the outer 
Solar System. The distribution of mass in this portion of the Solar System may provide clues to both 
the early nature of and the formation scenario for the Solar System. Current techniques for finding 
these objects that rely upon the detection of reflected sunlight are only sensitive to the largest 
and closest members of this population, objects that are rare and therefore are likely to be special 
cases. Our new techniques directly probe the low mass end of the distribution that is otherwise
quite difficult to detect.

It has been suggested that observations of occultation events may be used to constrain the low mass end 
of the TNO distribution. Bodies in the Kuiper Belt can be detected when they transit in front of distant 
luminous objects, resulting in a brief dip in the observed brightness. The {\it Taiwan-America Occultation 
Survey} (TAOS) \citep{Lehner2006} seeks to conduct a census of large Kuiper Belt objects using background 
stars. The cadence of observation required to detect the smallest objects presents significant technological 
challenges because of optical CCD read-noise. This same occultation method has been applied to Scorpius X-1, 
the brightest X-ray source near the ecliptic plane \citep{Chang06}, since X-ray detectors can be operated at 
a higher cadence than optical CCDs. However, the small number of bright X-ray sources near the ecliptic
limits the total volume of the Kuiper Belt that can be surveyed. 

Our method does not suffer from these limitations as the CMB detector technology is proven and the CMB can be
accurately observed over nearly the entire sky (except for the Galactic plane). The main drawback of our
proposed techniques is their reliance on the unknown TNO mass function. We must also be careful when interpreting 
our results because we are basically constraining the low mass end of the TNO distribution and then using 
those results to infer the total mass in the outer Solar System. If the slope of the high mass end of the
mass function is steep ($\beta > 2$), then most of the mass lies in the low mass end and this inference
is valid. However, if the slope is relatively shallow ($\beta < 2$) then we should only interpret our results
as pertaining to the mass in small objects. Current observational results allow for both possibilities \citep{Bernstein04}
and hopefully future observations will determine the true slope. These concerns are in addition to those expressed
in \S \ref{sec:mean} regarding the possibility that rare high mass objects could cause the true total mass to differ 
from that inferred by extrapolating the mass function. Fortunately, the large objects are the easiest to optically
detect and our proposed techniques can work together with these surveys to provide us with an accurate description
of the outer Solar System.

The next stage is to use the data provided by WMAP and analyze it as discussed in \S \ref{sec:WMAP} in
order to produce a map of Kuiper Belt objects. The data analysis, which must be done with the time-ordered
data, is a lengthy process, which is why it was not included in this paper.  The potential to better understand
the formation of our Solar System, and subsequently other planetary systems as well, will certainly justify the effort.
 
\acknowledgments 
The authors would like to thank an anonymous referee whose helpful comments improved the quality of this manuscript. We would like to thank C. Alcock, M. Brown, D. Finkbeiner, D. Fixsen, P. Goldreich, C. Hirata, M. Holman, S. Kenyon, 
A. Loeb,  L. Page, M. Pan, G. Rybicki, R. Sari, D. Spergel and P. Thaddeus for helpful conversations. 
DB thanks the hospitality of the 
Harvard Institute for Theory and Computation where some of this work was completed and acknowledges financial support 
from the Betty and Gordon Moore Foundation. CLS was supported under the National Science Foundation Graduate 
Research Fellowship Program. CHB acknowledges support from the Harvard Origins of Life Initiative.


\begin{thebibliography}{99xx}

\bibitem[Babich \& Loeb(2007)]{Babich06} Babich, D. \& Loeb, A.\ 2007, in preparation
\bibitem[Backman et al.(1995)]{Backman95}  Backman, D.~E., Dasgupta, A., \& Stencel, R.~E.\ 1995, \apjl, 450, L35 
\bibitem[Bernstein et al.(2004)]{Bernstein04}  Bernstein, G.M. et al. 2004, \aj, 128, 1364 
\bibitem[Brown et al.(2004)]{Brown04} Brown, M.~E., Trujillo, C.~A., \& Rabinowitz, D.~L. 2004, \apj, 617 645
\bibitem[Brown et al.(2005)]{Brown05} Brown, M.~E., Trujillo, C.~A., \& Rabinowitz, D.~L. 2005, \apjl, 635, L97 
\bibitem[Burns et al.(1979)]{Burns79} Burns, J.~A., Lamy, P.~L., \& Soter, S.\ 1979, Icarus, 40, 1 
\bibitem[Chang et al.(2006)]{Chang06} Chang, H.-K., et al. \ 2006, \nat, 442, 660
\bibitem[Draine(2003)]{Draine03} Draine, B.~T. \ 2003, \araa, 41, 241 
\bibitem[Dones et al.(2004)]{Dones04} Dones, L. et al. 2004, ASPC, 323, 371
\bibitem[Fernandez(1997)]{Fernandez97} Fernandez, J.~A.\ 1997, Icarus, 129, 106 
\bibitem[Fixsen et al.(1996)]{Fixsen96} Fixsen, D.~J., et al. \ 1996, \apj, 473, 576 
\bibitem[Fixsen et al.(1997)]{Fixsen97} Fixsen, D.~J., Hinshaw, G., Bennett, C.~L., \& Mather, J.~C. 1997, \apj, 486, 623 
\bibitem[Fixsen \& Dwek(2002)]{Fixsen02a} Fixsen, D.~J., \& Dwek, E.\ 2002, \apj, 578, 1009
\bibitem[Fixsen \& Kogut(2002)]{Fixsen02}Fixsen, D.~J. \& Kogut, A. 2002, \apj, 581, 817
\bibitem[Goldreich et al.(2004)]{Goldreich04} Goldreich, P., Lithwick, Y., \& Sari, R.\ 2004, \araa, 42, 549 
\bibitem[Greenberg(1978)]{Greenberg78} Greenberg, J.M., in McDonnell, J.A.M., ed., {\it Cosmic Dust}, Chichester, p. 187 (1978)
\bibitem[Hamid et al.(1968)]{Hamid68} Hamid, S.~E., Marsden, B.~G., \& Whipple, F.~L.\ 1968, \aj, 73, 727 
\bibitem[Heisler \& Tremaine(1986)]{Heisler86} Heisler, J. \& Tremaine, S. 1986, Icarus, 65, 13
\bibitem[Hills(1981)]{Hills81} Hills, J.~G.\ 1981, \aj, 86, 1730 
\bibitem[Hogg et al.(1991)]{Hogg91} Hogg, D.~W., Quinlan, G.~D., \& Tremaine, S.\ 1991, \aj, 101, 2274 
\bibitem[Kenyon \& Windhorst(2001)]{Kenyon01} Kenyon, S.~J., \& Windhorst, R.~A.\ 2001, \apjl, 547, L69 
\bibitem[Kogut et al.(2006)]{Kogut06} Kogut, A. et al 2006, astro-ph/0609373
\bibitem[Lehner et al.(2006)]{Lehner2006} Lehner, M.J. et al. 2006, Astr. Noch., 327, 814

\bibitem[Lissauer \& Stevenson(2005)]{Lissauer2005} Lissauer, J. \& Stevenson, D.\ 2006, in {\it Protostars and Planets V}, B. Reipurth, D. Jewitt, and K. Keil (eds.), University of Arizona Press, Sec.7-1

\bibitem[Luu \& Jewitt(2002)]{Luu02} Luu, J.~X., \& Jewitt, D.~C.\ 2002, \araa, 40, 63 
\bibitem[Marsden \& Sekanina(1971)]{Marsden71} Marsden, B.~G. \& Sekanina, Z. 1971 \aj, 76, 1135

\bibitem[Mather et al.(1992)]{mather1992} Mather, J.~C. et al. 1991, in \textit{After the first three minutes; Proceedings of the 1st Astrophysics Workshop, Univ. of Maryland}. S.~S. Hold, C.~L. Bennett, V. Trimble (eds.).  p.43

\bibitem[Mumma et al(1993)]{Mumma93} Mumma, M.~J., Weissman, P.~R. \& Stern, S.~A. 1993, in 
{\it Protostars and Planets III}, Lunine, J.I. \& Levy, E.H., eds., University of Arizona Press, 1177
\bibitem[Oort(1950)]{Oort50} Oort, J.~H. 1950, \bain, 11, 91
\bibitem[Pan \& Sari(2005)]{Pan05} Pan, M., \& Sari, R.\ 2005, Icarus, 173, 342 

\bibitem[Peebles(1993)]{Peebles93} Peebles, P.~J.~E. 1993, {\it Principles of Physical Cosmology}, Princeton University Press, p. 131

\bibitem[Rybicki \& Lightman(1979)]{Rybicki79} G.B. Rybicki, A.P. Lightman\ 1979, {\it Radiative Processes in Astrophysics}, Wiley-Interscience, p. 16
\bibitem[Spitzer(1978)]{Spitzer78} Spitzer, L.\ 1978, {\it Physical Processes in the Interstellar Medium}, Wiley-Interscience, p. 157 
\bibitem[Stern \& Weissman(2001)]{Stern2001}Stern, A. \& Weissman, P.\ 2001, \nat, 409, 6820, 589
\bibitem[Stern(2003)]{Stern2003}Stern, A.\ 2003, Nature 424, 6949, 639
\bibitem[Standish(2004)]{Standish2004}Standish, E.M.\ 2004, in \textit{Proceedings of IAU Colloquium 196}, Kurtz, D.W. (ed.), p.163
\bibitem[Tegmark \& Silk(1994)]{Tegmark95} Tegmark, M. \& Silk, J 1994, \apj, 423, 529
\bibitem[Tegmark \& Efstathiou(1996)]{Tegmark96} Tegmark, M. \& Efstathiou, G.\ 1996, \mnras, 281, 1297 
\bibitem[Tegmark et al.(1997)]{Tegmark97} Tegmark, M., Taylor, A.~N., \& Heavens, A.~F.\ 1997, \apj, 480, 22 
\bibitem[Teplitz et al.(1999)]{Teplitz99} Teplitz, V.~L. et al.\ 1999, \apj, 516, 425 

\end{thebibliography}
\end{document}